\documentclass[twocolumn,showpacs,superscriptaddress,nofootinbib]{revtex4-1}

\usepackage{dcolumn}
\usepackage{amsmath}

\def\nuc#1#2{\relax\ifmmode{}^{#1}{\protect\text{#2}}\else${}^{#1}$#2\fi}

\newcommand{\br}{{\vec r}}
\newcommand{\bR}{{\vec R}}
\newcommand{\bK}{{\vec K}}
\newcommand{\bk}{{\vec k}}
\newcommand{\pvec}[1]{\vec{#1}\mkern2mu\vphantom{#1}}

\newcommand{\be}{\begin{eqnarray}}
\newcommand{\ee}{\end{eqnarray}}

\makeatletter

\providecommand{\LyX}{L\kern-.1667em\lower.25em\hbox{Y}\kern-.125emX\@}

\usepackage{graphics}
\usepackage{comment}

\makeatother

\begin{document}

\title{A three-body model for the analysis of quasi-free scattering reactions in inverse kinematics}

\author{A.~M.\ Moro}
\email{moro@us.es}
\affiliation{Departamento de F\'{\i}sica At\'omica, Molecular y
Nuclear, Facultad de F\'{\i}sica, Universidad de Sevilla,
Apartado~1065, E-41080 Sevilla, Spain}

\date{\today}

\begin{abstract}
A new method to calculate cross sections for $(p,pn)$ and $(p,2p)$ reactions measured under inverse kinematics conditions is proposed. The method uses the prior form of the scattering transition amplitude, and replaces the exact three-body wave function appearing in this expression by an expansion in terms of $p$-$n$ or $p$-$p$ states, covering the physically relevant excitation energies and partial waves. A procedure of discretization, similar to that used in continuum-discretized coupled-channels calculations, is applied in order to make this expansion finite and numerically tractable. The proposed formalism is non-relativistic but several relativistic kinematical corrections are applied to extend its applicability to energies of current interest. The underlying optical potentials for the entrance and exit channels   are generated microscopically, by folding an effective density-dependent G-matrix with the density of the composite nucleus. Numerical calculations for $^{12}$C($p$,$2p$), $^{12}$C($p$,$pn$) and $^{23}$O($p$,$pn$) at $\sim$400~MeV/nucleon are presented to illustrate the method. The role of final-state interactions and Pauli principle between the outgoing nucleons is also discussed.

\end{abstract}


\maketitle

\vspace{2cm}
\section{Introduction}
Quasi-free scattering (QFS) experiments of the form $(p,pn)$  and $(p,2p)$ (hereafter $(p,pN)$) have been used extensively as a tool to extract spectroscopic information of proton-hole and neutron-hole states in nuclei, such as  separation energies, spin-parity assignments, and occupation probabilities. In these reactions, an energetic proton beam ($E > 100$~MeV) collides with a stable target nucleus, removing one or more nucleons, and leaving a residual nucleus,  either in its ground state, or in an excited state.   

Recently, the technique has been extended to the study of unstable nuclei, using inverse kinematics, i.e., bombarding a hydrogen target with a energetic radioactive beam. This technique is analogous to the knockout experiments with composite targets used extensively in the past years \cite{Nav98,Aum00,Mad01,Han03,Gad08}. Although both kinds of experiments are meant to provide similar information, there are important differences between them. In knockout reactions, a nucleon is suddenly removed from the fast-moving projectile after colliding with a light target nucleus, like $^{9}$Be. Due to the strongly absorptive nature of the core-target interaction at these energies, the process is highly peripheral and hence mainly dependent on the tail of the wave function of the removed nucleon or, more correctly, on the overlap function between the projectile and residual core wave functions. The norm of this overlap is the spectroscopic factor, a quantity that is directly related to the occupation probability of a given single-particle orbital. Since this norm depends on the full overlap, and not just on its tail, this raises the question of the reliability of the spectroscopic information extracted from a process that is only sensitive to a small piece of the wave function. On the other hand,   $(p,pN)$ reactions are expected to be more sensitive to deeper portions of the wave function and so they should help to reduce these ambiguities. Consequently, the information obtained from these experiments will be complementary to that obtained from heavy-ion knockout reactions at intermediate energies, and from transfer reactions, at lower energies. 

Some leading facilities, such as GSI, RIKEN and NSCL/MSU, have plans to perform inverse kinematics experiments with exotic beams. Consequently, it is of timely importance to revisit theoretical methods to analyse this kind of processes.  
  
Theoretical analyses of the ($p$, $pN$) reactions with stable nuclei have been commonly done using the distorted-wave impulse approximation (DWIA) \cite{Jac66a,*Jac66b}. Roughly speaking, the impulse approximation (IA) means that the binding potential of the removed particle can be neglected in comparison with the projectile--target kinetic energy (see, e.g.~\cite{Gol04}, Chap.~11). At the energies usually employed in QFS experiments (several hundreds of MeV per nucleon) this approximation is expected to be justified. The DWIA method is usually formulated in terms of the nucleon-nucleon transition matrix (T-matrix hereafter); the IA involves the replacement of this operator by a {\it free} T-matrix between the incident proton and the struck nucleon. Practical implementations of the DWIA formalism commonly involve further approximations, such as the substitution of   this T-matrix by its on-shell value, or its representation by a zero-range operator. If reliable structure information is to be extracted from these experiments, these approximations need to be revisited and tested. Consequently, the validity of the DWIA formalism should be investigated comparing with more elaborate reaction theories  \cite{Cres08,Cres09,Cres14}. 

A three-body reaction framework which does not make use of the IA is the continuum-discretized coupled-channels (CDCC) method \cite{Aus87}. This method has been very successful in the analysis of reactions induced by weakly-bound projectiles at low and medium energies. The method has been also applied to one-neutron removal reactions on proton targets at intermediate energies ($\sim$40-70 MeV/nucleon) \cite{Oza11,Kon09}. For a knockout reaction of the form $A(p,pN)C$, the standard CDCC method aims at expanding the three-body wave function of the system in terms of $N-C$ eigenstates. To make the expansion finite, the $N-C$ continuum must be truncated in energy and angular momentum and discretized.    However, the large angular momentum and energy transfer found in these reactions makes  that convergence of the observables requires a very large model space. For energies of current interest, of several hundreds of MeV/u, the method becomes unpractical.  

Benchmark calculations with the Faddeev/AGS method \cite{Del07} for the $^{11}$Be($p$,$pn$) reaction at $\sim$35~MeV/u showed that, while the CDCC expansion of the breakup states in terms of $n$-$^{10}$Be states converged very slowly with the size of the model space, the alternative expansion in terms of $p$-$n$ states required a much smaller model space to reproduce the dominant part of the $^{10}$Be inclusive cross sections. This alternative expansion makes use of the prior-form representation of  the transition amplitude, in which the three-body wave function is approximated by an expansion in a basis of $p$+$N$ states.  A procedure of continuum discretization, similar to that used in standard CDCC calculations, is used for the $p$-$N$ states. The resultant expression is formally  similar to the CCBA expression commonly used in transfer reactions and, therefore, in some previous applications \cite{Mor06a}, the method has been referred to as {\it transfer to the continuum} method. This allows its implementation in standard coupled-channels codes  after some suitable modifications.

In this paper, I present some exploratory calculations to illustrate the application of this method to the interpretation of  $(p,2p)$ and $(p,pn)$ experiments in inverse kinematics.  Although the general formulation of the method has been presented before  \cite{Mor06a,Del07}, it is described in more detail here and some suitable prescriptions for its input ingredients (internal wave functions, optical potentials, etc) as well as several relativistic kinematic corrections (to make the model applicable to higher energies) are discussed. The role of Pauli principle and final-state interactions of the outgoing nucleons is also discussed. The calculations here presented are for relatively high energies ($\sim$400~MeV/u) but, since the method does not rely on the IA, it might be applicable also to lower energies, for which the DWIA may not be adequate.




The paper is structured as follows. In Sec.~\ref{sec:theory} the theoretical formulation of the method is presented. This section discusses also the choice of the NN interaction, which is the main responsible for the $(p,pN)$ process,  the construction of appropriate optical potentials including in-medium effects, and some relativistic corrections applied to the model. In Sec.~\ref{sec:results} the method is applied to the $^{12}$C($p$,$pN$) and $^{23}$O($p$,$pn$) reactions at $E\sim400$ MeV/nucleon. In Sec.~\ref{sec:discus} the connection with the DWIA method is discussed. Finally, Sec.~\ref{sec:summary} summarizes the main results of this work.

\section{Theoretical model \label{sec:theory}}

\subsection{The transition amplitude}
\begin{figure}[t]
{\par\centering \resizebox*{0.45\textwidth}{!}
{\includegraphics{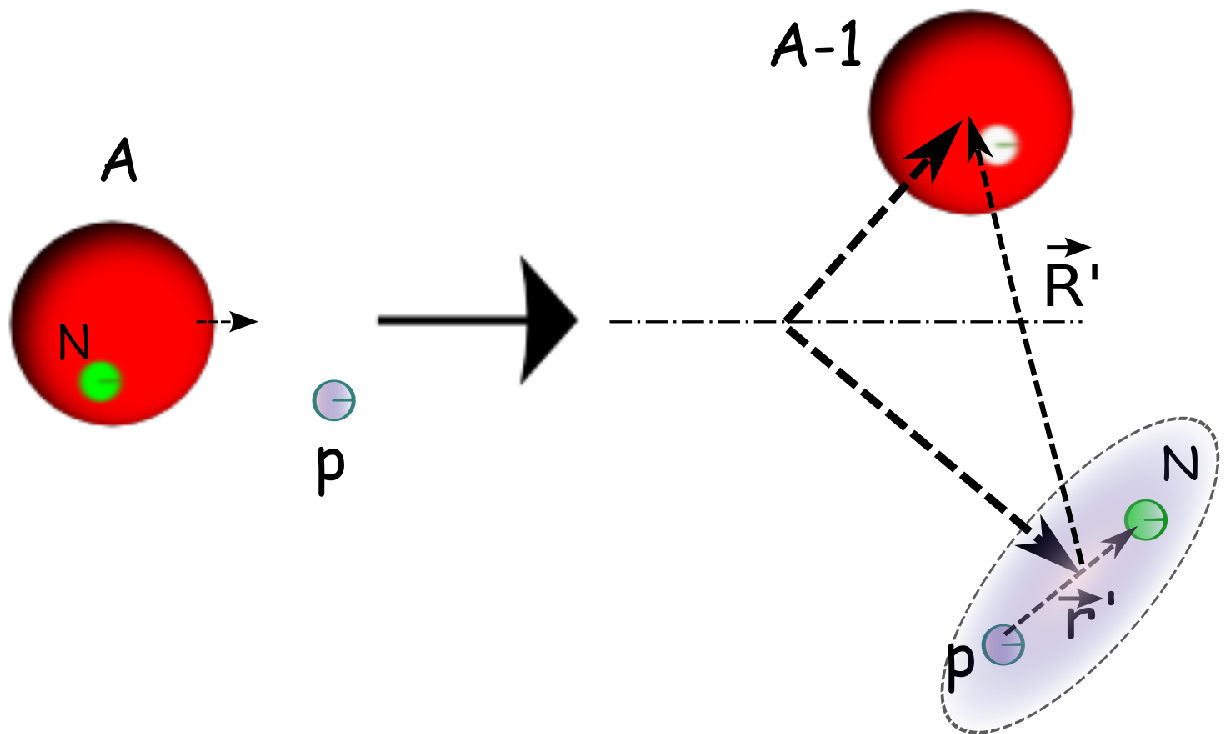}}\par}
\caption{\label{fig:qfs} (Color online) Diagram for a $(p,pN)$ reaction in inverse kinematics, modeled as a binary process. }
\end{figure}

Let us consider a reaction of the form 
\begin{equation}
A + p \rightarrow C(\alpha) + p + N ,
\end{equation}
 in which an incident composite nucleus $A=C+N$  collides with a proton target, losing a nucleon (proton or neutron) and giving rise to a residual core nucleus ($C$) in some definite state $\alpha$ and  two outgoing nucleons ($p+n$ or $p+p$). The process is schematically depicted in Fig.~\ref{fig:qfs}. 

Using the prior form of the transition amplitude, the exact transition amplitude for this reaction can be written as
\begin{equation}
{\cal T}_{if}(\alpha) = \langle \Psi^{(-)}_{\alpha,f} |  V_{pN} + V_{pC} |\phi_A(\xi_A) e^{i \bK_{pA} \bR} \rangle  ,
\label{Texact}
\end{equation}
where $\phi_A(\xi_A)$ represents the ground state wave function of the nucleus $A$, with $\xi_A$ denoting its internal coordinates. The plane wave $e^{i \bK_{pA} \bR}$ describes the relative motion of the $p+A$ system. 
At this stage, the potential $V_{pC}$ is a many-body operator containing the interaction of the target proton with all the core nucleons. The final function $\Psi^{(-)}_{\alpha,f}$ is the exact scattering wave function subject to the boundary conditions consisting of a plane wave (or Coulomb wave) in the channel $f$ (corresponding to some definite state for the relative motion of the three outgoing fragments), with the core in state $\alpha$, and ingoing spherical waves in all other open channels. This wave function is a solution of  the following many-body Schr\"odinger equation
\begin{align}
\label{eq:Psi_exact}
[E^- - K_{r'}- K_{R'}  - V_{pN}-V^\dagger_{pC} - V^\dagger_{NC} ] \Psi^{(-)}_{\alpha,f} (\pvec{r}',\pvec{R}',\xi_C)&=0 ,
\end{align}
where $E^-= E-i\epsilon$ and $\xi_C$ denotes the core internal coordinates.  Note that  the final three-body state has been expressed in terms of the Jacobi coordinates $\{ \pvec{r}',\pvec{R}' \}$ (see Fig.~\ref{fig:qfs}). 

Equation (\ref{Texact}) gives the exact transition amplitude of the many-body scattering problem. It  can be formally reduced to an effective three-body problem, using the following approximation for $\Psi^{(-)}_{\alpha,f}$:
\begin{equation}
\Psi^{(-)}_{\alpha,f}(\pvec{r}',\pvec{R}',\xi_C) \simeq  \Psi^{3b(-)}_{f}(\pvec{r}',\pvec{R}') \phi^\alpha_C(\xi_C) ,
\end{equation}
where $\phi^\alpha_C(\xi_C)$ is the core wave function in the state $\alpha$ and $\Psi^{3b(-)}_{f}(\pvec{r}',\pvec{R}')$ is a three-body wave function obtained as a solution of the following effective Schr\"odinger equation:
\begin{align}
\label{eqPsi}
[E^- - K_{r'}- K_{R'}  - V_{pN}-U^\dagger_{pC} - U^\dagger_{NC} ] \Psi^{3b(-)}_{f} (\pvec{r}',\pvec{R}')&=0 ,
\end{align}
where $U_{pC}$ and  $U_{NC}$ are now effective nucleon-nucleus interactions which, in practice, will be replaced by optical model potentials at the apropriate energy per nucleon. 

In this three-body model, the transition amplitude is 
\begin{equation}
{\cal T}^{3b}_{if}(\alpha) = \langle \Psi^{3b(-)}_{f} \phi^\alpha_C(\xi_C) |  V_{pN} + U_{pC} |\phi_A(\xi_A) e^{i \bK_{pA} \bR} \rangle  .
\label{T3b}
\end{equation}
If the potential $U_{pC}$ is taken to be independent of the internal  coordinates of $C$ ($\xi_C$), as it is usually assumed, one can perform the integral over these internal coordinates, to give
\begin{equation}
\label{eq:overlap}
\int d\xi_C \phi^\alpha_C(\xi_C) \phi_A(\xi_A) = \sqrt{S_{\alpha,\ell,j}}\varphi^\alpha_{CA}(\br) ,
\end{equation}
where $ \sqrt{S_{\alpha,\ell,j}} \phi^\alpha_{CA}(\br)$ is an overlap wave function, with $\phi^\alpha_{CA}(\br)$ a unit normalized wave function depending on the relative coordinate of the removed particle with respect to the core and $S_{\alpha,\ell,j}$  the spectroscopic factor.  Using (\ref{eq:overlap}) in Eq.~(\ref{T3b}) one finds,
\begin{equation}
{\cal T}^{3b}_{if}(\alpha) =  \sqrt{S_{\alpha,\ell,j}} \langle \Psi^{3b(-)}_{f} |  V_{pN} + U_{pC} |\varphi^\alpha_{CA} e^{i \bK_{pA} \bR} \rangle  .
\label{T3b-bis}
\end{equation}

As done in transfer calculations, it is convenient to introduce an auxiliary potential in the incoming channel, $U_{pA}(\bR)$, so the previous equation transforms to:
\begin{equation}
{\cal T}^{3b}_{if}(\alpha) =  \sqrt{S_{\alpha,\ell,j}} \langle \Psi^{3b(-)}_{f} |  V_{pN} + U_{pC} - U_{pA} |\varphi^\alpha_{CA} \chi^{(+)}_{pA} \rangle ,  
\label{T3b_uaux}
\end{equation}
where $\chi^{(+)}_{pA}$ is the distorted wave generated by the potential $U_{pA}(\bR)$. Note that, when $\Psi^{3b}_{f}$ is the exact solution of 
(\ref{eqPsi}), the amplitude (\ref{T3b_uaux}) is strictly independent of the choice of $U_{pA}$. In practical calculations, in which $\Psi^{3b}_{f}$ must be approximated somehow, the result will however depend on this potential. The usual choice is to use for $U_{pA}$ an optical potential describing the elastic scattering of the $p+A$ system. With this choice, one expects that the difference $U_{pC} - U_{pA}$ (the so-called {\it remnant term}) will contribute little to the integral and hence the matrix element will be mostly determined by the $V_{pN}$ interaction.

In order to reduce Eq.~(\ref{eqPsi}) to a tractable form,  $\Psi^{3b(-)}$ is expanded in terms of $p+N$ eigenstates, i.e.,
\begin{align}
\label{eq:Psi_cont}
\Psi^{3b(-)}_{f} (\pvec{r}',\pvec{R}') = \sum_{j^\pi} \int dk ~\phi^{j\pi}(k,\pvec{r}') \chi_{j,\pi}(\bK,\pvec{R}')  ,
\end{align}
where $\bk$ is the relative wave number of the  $p$+$N$ pair, $\bK$ that for the relative motion between the residual nucleus $C$ and the $pN$ pair and  $\chi_{j,\pi}(\bK,\pvec{R}')$ is a function describing the relative motion of the $p$+$N$ system with respect to the residual nucleus, when the former is in a given final state $\{k,j^\pi\}$. 
[Note that, in actual calculations, this expansion will be done in terms of states of total angular momentum $J$. To simplify the discussion, the notation used here is somewhat schematic].
%
%
%
 The states $\phi^{j\pi}(k,\pvec{r}')$ are the eigenstates of the $p$+$N$ Hamiltonian using the potential $V_{pN}(\pvec{r}')$. In the $(p,pn)$ case, the expansion (\ref{eq:Psi_cont}) will contain also a term for the deuteron ground state. This term is omitted for simplicity of the notation. In the 
 $(p,2p)$ case, the $p+p$ wave function must be antisymmetrized to account for the indistinguishability  of the outgoing protons. In practice, this restricts the allowed values of $j^\pi$ and introduces a factor of 2 in the calculated cross sections with respect to the unsymmetrized calculation.%
\footnote{Note that, treating the  $(p,2p)$ reaction as a binary process of the form  $p+A \rightarrow$ $^2${He}$ + C$, the 2 factor can be regarded as the spectroscopic factor for the overlap $\langle ^2{\rm He}|p \rangle$.}
Energy conservation in the final channel (non-relativistic by now) implies that
\begin{align}
\label{eq:etot}
E= e_{pN}+ E_\mathrm{cm}=\frac{\hbar^2 k^2}{2 \mu_{pN}} + \frac{\hbar^2 K^2}{2 \mu_{pN,C}},
\end{align}
where $E$ is the total energy of the system, $e_{pN}$ the relative energy of the $p+N$ pair ($e_{pN}=-2.22$~MeV for the deuteron ground state), and  $E_\mathrm{cm}$ the kinetic energy associated with the relative motion of the core with respect to the CM of the $p+N$ pair. 

The  functions $\chi_{j,\pi}(\bK,\pvec{R}')$ could  in principle be obtained by inserting the expansion (\ref{eq:Psi_cont}) into Eq.~(\ref{eqPsi}). However, since these functions depend upon the continuous parameter $K$, this would give rise to an infinite number of equations. In practice, one may use a discretization procedure similar to that used in the Continuum-Discretized Coupled-Channels method \cite{Aus87}, in which the final $p+N$ states are grouped ({\it binned}) in energy or momentum intervals as
\begin{align}
\label{PhiCDCC}
\Psi^{3b(-)}_{f} \approx \Psi^\mathrm{CDCC}_{f}=\sum_{n,j,\pi}  \phi^{j\pi}_{n}(k_{n},\pvec{r}') \chi_{n,j,\pi}(\bK_{n},\pvec{R}') ,
\end{align}
%
%
%
where $k_n$ are some  average values for the discretized $p$-$N$ energies and  $\phi^{j\pi}_{n}(k_n,\pvec{r}')$ the bin wave functions. Note that the subscript $f$ in $\Psi^\mathrm{CDCC}_{f}$ retains the information on the final state, corresponding to some definite values of $n$, $j$ and $\pi$. Details on the construction of these bins can be found elsewhere \cite{Aus87,TN09}.

The angular differential cross section for a given final discretized bin $f = \{ n,j, \pi \}$, and a given core state $\alpha$ is 
\begin{align}
\label{eq:dsdw_bin}
 \frac{d\sigma_{n,j,\pi}(\alpha)}{d\Omega_c}  & =   \frac{1}{(2s_p+1)(2J_A+1)} \nonumber \\
 &\times \frac{\mu_i \mu_f}{(2 \pi \hbar^2)^2} \frac{K_n}{K_i} \sum_{\sigma}|{\cal T}^{3b}_{i,f}(\alpha) |^2 ,
\end{align}
with $\mu_{i}$ ($\mu_{j}$) the reduced masses of the initial (final) mass partition and ${\cal T}^{3b}_{i,f}(\alpha)$ the transition amplitude obtained by replacing the CDCC expansion (\ref{PhiCDCC}) in the transition amplitude (\ref{T3b_uaux}). The sum in $\sigma$ includes the spin projections of the outgoing $pN$ pair and of the residual nucleus $C$. The angle specified by $\Omega_c$ is the scattering angle of the core in the center-of-mass (CM) frame. 

The double differential cross section, with respect to the scattering angle of the core  and the internal energy of the $p$+$N$ system, can be obtained at the discretized energies $e_{pN}= e^n_{pN}$ as
\begin{align}
\label{eq:dsdw}
\left . \frac{d^2\sigma_{j,\pi}(\alpha)}{d e_{pN} d\Omega_c} \right |_{e_{pN}=
 e^n_{pN} } \simeq  \frac{1}{\Delta_n} \frac{d\sigma_{n,j,\pi}(\alpha)}{d\Omega_c} ,
\end{align}
where $\Delta_n$ is the width of the bin to which the energy $e_{pN}$ belongs. This can be readily transformed to a double differential cross section with respect to the energy of the outgoing core in the overall CM frame ($E_c$), using the usual non-relativistic relation for binary collisions
\begin{align}
\label{eq:ec}
E_c = \frac{m^{*}_{pN}}{m^{*}_{pN} + m_c} E_\mathrm{cm} ,
\end{align}
where $m^{*}_{pN}= m_{p}+m_{N} +e_{pN}$. Thus, 
\begin{align}
\label{eq:dsdwc}
\frac{d^2\sigma_{j,\pi}(\alpha)}{d E_c d\Omega_c} =
\frac{m^{*}_{pN}}{m_c + m^{*}_{pN}} \frac{d^2\sigma_{j,\pi}(\alpha)}{d e_{pN}  d\Omega_c} .
\end{align}

The inclusive cross section will be then obtained summing the contributions from all final $j^\pi$ configurations
\begin{align}
\label{eq:dsdwc-inc}
 \frac{d^2\sigma(\alpha)}{dE_c d\Omega_c}  =  \sum_{j,\pi} \frac{d^2\sigma_{j,\pi}(\alpha)}{d E_c d\Omega_c} .
\end{align}

Note that, Eq.~(\ref{eq:ec}), along with Eq.~(\ref{eq:etot}), establishes a one-to-one correspondence between the energy of the core, in the CM frame, and the internal energy of the $p$+$N$ pair. Note also that Eq.~(\ref{T3b_uaux}) resembles the transition amplitude for a transfer process, analogous to that appearing in the standard CCBA method for binary collisions \cite{Sat83}. Taking advantage of this formal analogy, one can evaluate this transition amplitude using standard coupled-channels codes. For the calculations presented in this work, the code {\sc fresco} \cite{fresco} has been used, with some suitable modifications described below.

\subsection{Relativistic kinematics \label{sec:relkin}}
Ongoing and planned QFS experiments in inverse kinematics are performed at energies of several hundreds of MeV per nucleon. At these energies, relativistic kinematics is clearly important. Although the treatment presented in the previous sections is non-relativistic, some relativistic corrections can be readily implemented in order to take into account, at least approximately, these effects.

The total energies of the incident nucleus and the proton target in the center-of-momentum  frame are given 
by ($c=1$ in this section)
\begin{align}
\varepsilon_A = \frac{s+(m_A -m_p)}{2 \sqrt{s}} ; \quad
\varepsilon_p = \frac{s-(m_A -m_p)}{2 \sqrt{s}} ,
\end{align}
where $s$ is the usual Mandelstam invariant, corresponding to the square of the total energy. Assuming that the proton target is initially at rest, 
\begin{align}
s= (m_p + m_A)^2 + 2 m_p T_\mathrm{LAB} ,
\end{align}
where $T_\mathrm{LAB}$ is the kinetic energy of the projectile. 

Analogously, for the exit channel, the total relativistic energy of the outgoing core  can written as 
\begin{align}
\label{eq:eps_core}
\varepsilon_c = \frac{s+(m_c -m^{*}_{pN})}{2 \sqrt{s}} 
\end{align}
where one has exploited again the analogy of the process under study with a binary process of the form $A+p \rightarrow C + (pN)$.
Since $m^{*}_{pN}=m_p+m_N+e_{pN}$, and  $E_c=\varepsilon_c-m_c$, Eq.~(\ref{eq:eps_core}) relates the relative energy of the outgoing $p$-$N$ pair to the kinetic energy of the core. Using this relation, the generalization of the double differential cross section, Eq.~(\ref{eq:dsdwc}), results
\begin{align}
\label{eq:dsdwc_rel}
\frac{d^2\sigma_{j,\pi}(\alpha)}{d E_c d\Omega_c} =
\frac{\sqrt{s}}{m^{*}_{pN}} \frac{d^2\sigma_{j,\pi}(\alpha)}{d e_{pN}  d\Omega_c} ,
\end{align}
which reduces to Eq.~(\ref{eq:dsdwc}) in the non-relativistic limit. 

The modification of the distorted waves due to relativistic kinematics has been also studied. Unfortunately, there is no  unique prescription to incorporate relativistic kinematics within the Schr\"odinger scheme. Two common prescriptions used in the context of nucleon-nucleus scattering at intermediate energies have been considered here. The first one (see e.g.~\cite{Ray92}) consists in replacing  the reduced mass appearing in the kinetic energy term of the  Schr\"odinger equation (\ref{eqPsi}) by the so-called {\it reduced energy}. For example, for the incident channel,
\begin{align}
\mu_i \equiv \mu_{pA} \rightarrow \varepsilon_{pA} = \frac{\varepsilon_{p} \varepsilon_{A}}{\varepsilon_{p}+  \varepsilon_{A}} .
\end{align}
Inserting this definition into the Schr\"odinger equation, and multiplying the full equation 
by $\varepsilon_{pA}/\mu_{pA}$, one finds an equivalent Schr\"odinger equation with the usual reduced mass in the kinetic energy operator, but with the potential scaled by the factor 
\begin{equation}
\label{eq:gam1}
\gamma=\varepsilon_{pA}/\mu_{pA}.
\end{equation}

The second prescription considered here (see e.g.~\cite{Nad81}) assumes that the motion of the heavy nucleus in the CM frame can be treated nonrelativistically, whereas the light particle (the proton) is treated relativistically. A relativistic Schr\"odinger-type equation is then generated by reducing the Dirac equation for a massive energetic fermion of mass $m_p$ and relativistic wave number $k_p$, moving  in a central potential $U(r)$, verifying $U \ll m_p$ and $\nabla U \ll k_p$, both well satisfied for intermediate-energy proton scattering. In the reduced two-body problem with relativistic projectile but non-relativistic target the larger component of the wave function verifies a Schr\"odinger-like equation with a potential renormalized by the factor
\be
\label{eq:gam2}
\gamma= \frac{2 (E-M_A)}{E-M_A+m_p} .
\ee
Note that, in both prescriptions, the momentum to be used is the relativistic one. 

Finally, the expression for the differential cross section (\ref{eq:dsdw_bin}) is replaced by its relativistic counterpart (see e.g.~\cite{Joa75})
\begin{align}
\label{eq:dsdw_rel}
\frac{d\sigma_{n,j,\pi}(\alpha)}{d\Omega_c} & =   \frac{1}{(2s_p+1)(2J_A+1)} \frac{1}{(2\pi)^2 \hbar v_i}  \nonumber \\
  &\times 
 \frac{K_f}{\hbar^2 c^2  (1/\varepsilon_c + 1/\varepsilon_{pN})   } \sum_{\sigma} | {\cal T}^{3b}_{i,f}(\alpha) |^2  
\end{align}
where $v_i$ is the initial projectile-target relative velocity.
Experimental data usually consist of transverse or longitudinal momentum distributions of the residual core. These are readily obtained from the previous differential cross sections by applying the transformation
\begin{align}
\frac{d^2\sigma}{dE_c d\Omega_c} dE_c d\Omega_c & = \frac{d^3\sigma}{dp^3} d^3 \vec{p} 
\end{align}
giving rise to
\begin{equation}
\frac{d^3\sigma}{d p^3} = \frac{c^2}{p} \frac{1}{\sqrt{(pc)^2  + (mc^2)^2} } \frac{d\sigma}{dE_c d\Omega_c} .
\end{equation}
Decomposing the momentum in its transverse ($p_t$) and longitudinal ($p_z$) components, the differential volume becomes $d^3 \vec{p} = p_t dp_t d\phi dp_z$ (with $\phi$ the azimutal angle), and
\begin{equation}
\label{eq:dsdpt}
\frac{d\sigma}{dp_t} =  \pi p_t \int^{\infty}_{\infty} \frac{d^3\sigma}{d p^3} dp_z ,
\end{equation}
\begin{equation}
\label{eq:dsdpz}
\frac{d\sigma}{dp_z} = 2\pi \int \frac{d^3\sigma}{d p^3} p_t dp_t  .
\end{equation}
Writing $p_x= p_t \cos(\phi)$ and $p_y=p_t \sin(\phi)$ the transverse momentum distribution can be also decomposed into its $x$ and $y$ components, resulting,
\begin{equation}
\label{eq:dsdpx}
\frac{d\sigma}{dp_x} = \frac{1}{2\pi p_t} \int_{-\infty}^{\infty}  \frac{d\sigma}{d p_t} d p_y ,
\end{equation}
and likewise for ${d\sigma}/{dp_y}$.

\subsection{Nucleon-nucleon and nucleon-nucleus interactions \label{sec:omp}}
In the present approach, the nucleon-nucleon (NN) interaction appears in the transition operator [Eq.~(\ref{Texact})] as well as in the equation for the three-body (CDCC) wave function (\ref{eqPsi}). For this interaction, the Reid93 parametrization \cite{Reid93},  an updated regularized version of the pioneering Reid soft-core potential \cite{Reid68} developed by the Nijmejen group, has been adopted. This potential contains central, spin-orbit and tensor components, and reproduces accurately the proton-proton and proton-neutron phase-shifts up to an energy of 350 MeV ($\chi^2/N_\mathrm{data}=1.03$). 

The nucleon-nucleus potentials are calculated microscopically, by folding an effective NN interaction with the  nucleus ground-state density. Systematic studies of nucleon-nucleus  scattering at intermediate energies show that this procedure provides a good description of the elastic and inelastic data, provided that the effective NN interaction contains some energy and density dependence. Although $q$-space folding is usually preferable in folding calculations, the presence of this density dependence   makes  the use of $r$-space folding necessary which, for the central part, reads
%
%
\begin{equation}
\label{eq:rfold}
U(r)= \int d\pvec{r}' 
\left [ \rho_\mathrm{gs}(\pvec{r}') t^D(s,\rho_\mathrm{gs}) + j_0(k s) t^X(s,\rho) \rho(\br,\pvec{r}') \right ] ,
\end{equation}
where $t^D$ and $t^X$ correspond to the direct and exchange terms of the effective NN interaction, $k$ is the local wave number,  $\br$ and $\pvec{r}'$ are the projectile (nucleon) and target (nucleus) positions, $s= |\br-\pvec{r}'|$  and $\rho_\mathrm{gs}$ is the ground-state density, evaluated at the midpoint position $|\br-\pvec{r}'|/2$.

 The mixed transition density is represented by
\begin{equation}
\rho(\br,\pvec{r}')= \rho_\mathrm{gs}(\pvec{r}') C(k_F s),
\end{equation}
where $C$ is a correlation function describing the exchange non-locality and $k_F$ the Fermi momentum.   These folding calculations have been performed with the code {\sc lea} \cite{lea}. In the calculations presented in this work, the correlation function has been taken as unity, and the ground state densities are obtained from Hartree-Fock (HF) calculations.  The latter were computed with the code {\sc oxbash} \cite{oxbash}, using  the Skyrme Sk20 interaction. It is worth noting that these Skyrme HF densities reproduce well the nuclear radii extracted from electron scattering \cite{Ric03} and are of common use in the analysis of knockout experiments (see e.g.~\cite{Han03,Gad08}).


\subsection{Bound state wave functions  \label{sec:bs}}
According to Eq.~(\ref{eq:overlap}), the transition amplitude contains the overlap function between the initial ($A$) and final ($C$) nuclei. As it is usually done in the analysis of transfer and knockout reactions, this overlap function has been approximated by the single-particle wave function obtained as the solution of  a Woods-Saxon potential, with the depth adjusted to  reproduce the effective separation energy for the final state $\alpha$ of the core, namely, $S^*_{\alpha}= S_{n,p} + E_\alpha$, where  $S_{n,p}$ is the ground-state to ground-state nucleon separation energy and $E_\alpha$ the excitation energy of the core. Following \cite{Gad08}, 
the diffuseness parameter of the potential well was fixed at $a_0=0.7$~fm and the radius parameter $r_0$  was adjusted in order the rms of the calculated orbital coincide with that obtained from a HF calculation, i.e., $\sqrt{\langle r^2 _\mathrm{sp} \rangle}= [A/(A-1)]^{1/2} \sqrt{\langle r^2 _\mathrm{HF} \rangle}$.   Since the separation energy predicted by the HF calculation does not necessarily coincides with the experimental one, the HF separation energy is used for this fit. 


\section{Results \label{sec:results} }
To illustrate the method, some calculations for the reactions $^{12}$C($p$,$pN$) and    $^{22}$O($p$,$pn$)  at $\sim$400 MeV per nucleon are presented and discussed in this section. These calculations are performed with a modified version of the code {\sc fresco} \cite{fresco}, which incorporates the Reid93 NN interaction, and the relativistic kinematics corrections discussed in Sec.~\ref{sec:relkin}. 

Although Eq.~(\ref{T3b_uaux}), with the discrete expansion (\ref{PhiCDCC}) for the final three-body wave function, provides a numerically tractable form, the calculations can be significantly simplified, with a small loss of accuracy, making use of the additional approximation of neglecting couplings among $p$-$N$ continuum states with different $j^\pi$. The omission of these couplings will alter to some extent the distribution of flux between the final states. However, for an inclusive situation, in which the contribution from all the included $j^\pi$ must be added together, this redistribution of flux is not expected to affect significantly the core observables. 
 As an additional simplification,  the spin-orbit part of the nucleon-nucleus potentials in Eq.~(\ref{T3b_uaux}) has been neglected. 

\subsection{Application to $^{12}$C(\protect{$p$},\protect{$pN$})}
The $^{12}$C($p$,$2p$)$^{11}$B reaction at 400 MeV/nucleon, which coincides with the actual energy used in a recent experiment performed at GSI for this reaction  \cite{Tay11}, is consdered first. Only those processes leading to bound states of $^{11}$B are considered, which correspond to the removal of protons from the $1p_{3/2}$ and $1p_{1/2}$ orbitals, since  removal from the deeper $1s_{1/2}$ orbital will lead to unbound states of $^{11}$B. 

The required nucleon-nucleus optical potentials  are
$p$+$^{12}$C, in the incident channel, and $p$+$^{11}$B, in the exit channel. These potentials were generated with the microscopic folding approach described in Sec.~\ref{sec:omp}, using the  Paris-Hamburg (PH)  $G$-matrix  effective interaction \cite{Ger83,Rik84}. This interaction is energy and density dependent.    The $p$+$^{12}$C potential was calculated using the incident energy ($E_p= 400$~MeV/nucleon). For the exit channel, the choice is less clear, because the outgoing protons will emerge with a broad range of energies. For a pure QFS collision, one expects an average value of about $E_\mathrm{lab}/2$ for each nucleon, and so the outgoing optical potentials were evaluated at $E_p=200$~MeV in the present case. The ground-state densities of the $^{12}$C  and $^{11}$B nuclei were obtained from a Hartree-Fock (HF) calculation, using the Skyrme Sk20 interaction. To test the sensitivity of the calculated potentials with respect to this input, a two-parameter Fermi (2pF) parametrization of the  $^{12}$C density, extracted from electron scattering \cite{Far85}, was also considered.

\begin{figure}[t]
{\par\centering \resizebox*{0.45\textwidth}{!}
{\includegraphics{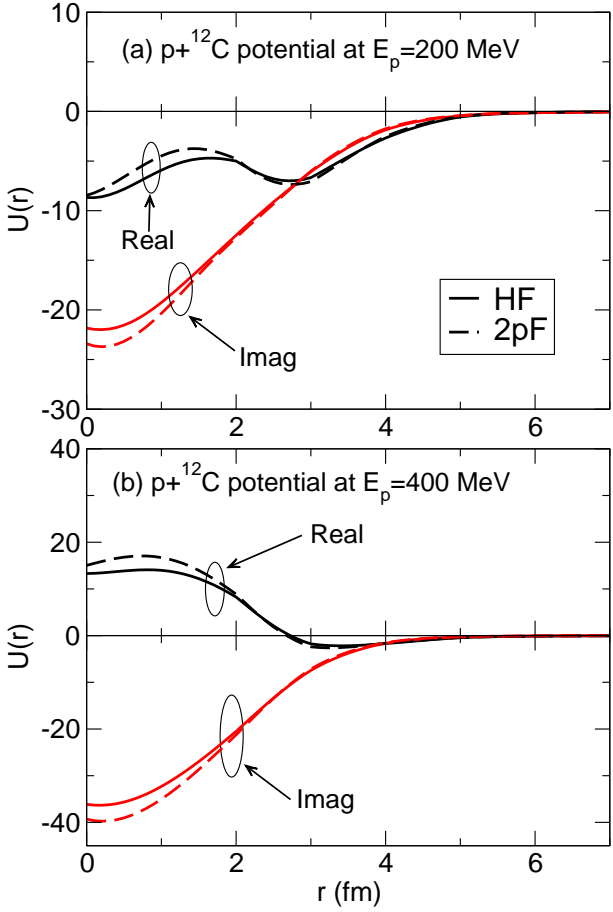}}\par}
\caption{\label{c12p_opt} (Color online) $p+^{12}$C microscopic optical potential generated with the Paris-Hamburg density-dependent NN interaction \cite{Ger83,Rik84},  folded with the ground-state density of $^{12}$C, at $E_p$=200 MeV (top) and 400 MeV (bottom). Solid and dashed lines correspond, respectively, to HF  (Sk20 interaction) and empirical densities obtained from electron scattering (the latter parametrized with a 2pF distribution).}
\end{figure}

Figure \ref{c12p_opt} shows the real and imaginary parts of the $p$+$^{12}$C central potential at 200 (upper) and 400 MeV (bottom). The solid and dashed lines are for the HF and 2pF densities.  It is interesting to note how, for the higher energy,  the potential becomes highly absorptive  and the real part, which is mainly attractive at low energies, develops a strong repulsive core.  The 2pF density gives a qualitatively similar behaviour. The calculated potentials show some differences at small distances ($<2$~fm) but are very similar at larger distances. Although $(p,pN)$ reactions are expected to explore shorter distances as compared to knockout reactions, they have still a peripheral nature. Therefore, one  expects that both potentials give similar $(p,pN)$ cross sections.

\begin{figure}[t]
{\par\centering \resizebox*{0.45\textwidth}{!}
{\includegraphics{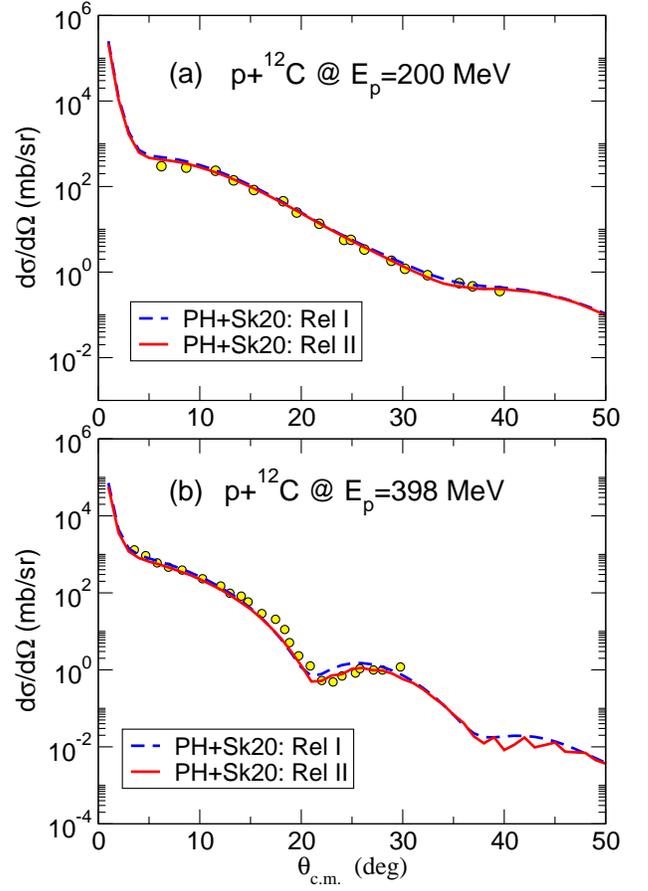}}\par}
\caption{\label{c12p_el} (Color online) Differential elastic cross section for $p$+$^{12}$C at 200 MeV (upper) and 398 MeV (bottom), using a  microscopic folding potential generated with the Paris-Hamburg effective NN interaction and the Skyrme HF density.  The dashed and solid lines correspond to the relativistic scaling prescriptions given by Eqs.~(\ref{eq:gam1}) and (\ref{eq:gam2}), respectively. The data are from Ref.~\cite{Jon86}.}
\end{figure}

The quality of the calculated potentials has been assessed comparing the calculated differential elastic cross section for 
or $p+^{12}$C at 200 and 398~MeV with the experimental data from Ref.~\cite{Jon86}.
For these calculations,  the folding potentials computed with the HF densities have been used. 
 The comparison  is shown in Fig.~\ref{c12p_el}, where the solid and dashed lines correspond to the relativistic scaling factors of Eq.~(\ref{eq:gam1}) (prescription I hereafter) and Eq.~(\ref{eq:gam2}) (prescription II), respectively. Both prescriptions reproduce fairly well the data at both energies, with prescription II providing a slightly better agreement. Therefore, the latter has been also adopted for the $(p,pN)$ calculations presented below. 

For the $(p,2p)$ calculations, the transition amplitude (\ref{T3b_uaux}) is evaluated with the CDCC expansion of the final three-body wave function  in terms of $p$-$p$ bins [Eq.~(\ref{PhiCDCC})]. For this expansion, 
partial waves $j \leq 5$ and relative energies ($e_{pp}$) up to the maximum allowed by energy conservation were used. Note that, in the $(p,2p)$ case,  $T=1$ for the outgoing pair and hence the generalized Pauli principle restricts the allowed final configurations to $\ell+S$=even, where $\ell$ is the orbital angular momentum for the $p$-$p$ pair and $S$ its total spin ($\vec{S}=\vec{s}_1 + \vec{s}_2$). This excludes some $j^\pi$ configurations, such as $1^+$ and $3^+$, which are purely $T=0$.   The bound-state wave function of the struck nucleon in the projectile  is generated with a Woods-Saxon potential, with diffuseness $a=0.70$~fm and the radius adjusted in order to reproduce the rms predicted by a HF calculation, as explained in Sec.~\ref{sec:bs}. This procedure yields the potential radius $R_0=2.953$~fm. Finally, the potential depth is adjusted to reproduce the ground-state to ground-state proton separation energy ($S_p=15.96$~MeV). 

\begin{figure}[t]
{\par\centering \resizebox*{0.45\textwidth}{!}
{\includegraphics{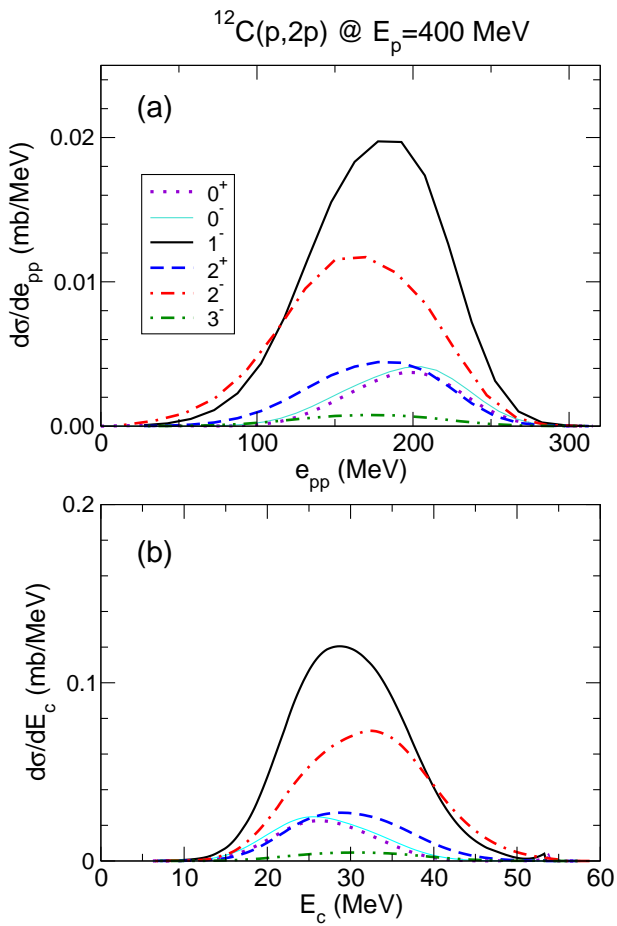}}\par}
\caption{\label{c12p2p_dsdex} (Color online) Top: relative energy distribution of the outgoing protons, following the process $^{12}$C($p$,$2p$), for several $j^\pi$ configurations of the $p$-$p$ pair. Bottom: corresponding energy distribution of the core, in the CM frame, calculated from Eq.~(\ref{eq:dsdwc_rel}). The calculations are done with microscopic distorting potentials (PH + Sk20), the relativistic prescription II and unit spectroscopic factor for the $^{12}$C$\rightarrow$$^{11}$B+$p$ decomposition. See text for details.}
\end{figure}

Before presenting the $(p,2p)$ cross sections,   the correspondence between the relative energy of the outgoing $p$-$p$ pair and the kinetic energy of the core in the CM frame,  given by Eqs.~(\ref{eq:etot}) and (\ref{eq:ec})], is discussed. This is illustrated  in Fig.~\ref{c12p2p_dsdex}. The upper panel shows the contribution of the dominant $j^\pi$ $p$-$p$ configurations, as a function of the relative energy $e_{pp}$. For simplicity, the spectroscopic factor has been set equal to unity. It is seen that a wide range of energies, from  $e_{pp}=0$ to  $e_{pp} \simeq 300$~MeV, are populated, with a maximum around $\sim$180~MeV. It is also seen that most of the cross section comes from the $j<3$ configurations, with $j=1^-$ giving the dominant contribution. It is therefore very important that the underlying NN interaction used in the evaluation of the scattering amplitude reproduces correctly at least these partial waves.   In the bottom panel, the same distributions are plotted as a function of the CM core energy, according to the relation (\ref{eq:eps_core}). Although this expression is  relativistic, it is clear from  this figure that the core energies, in the overall CM frame, can be treated nonrelativistically, as assumed in the relativistic prescription II discussed in the previous section.   

\begin{figure}[t]
{\par\centering \resizebox*{0.45\textwidth}{!}
{\includegraphics{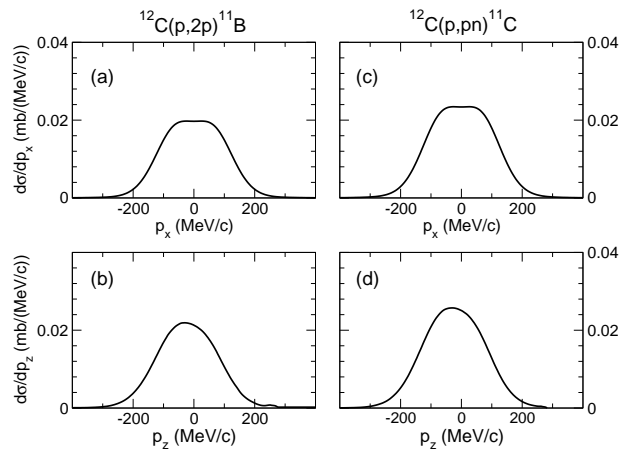}}\par}
\caption{\label{c12p_dsdpx} Transverse (top) and longitudinal (bottom) momentum of the residual core nucleus for the $^{12}$C($p$,$2p$) (left) and $^{12}$C($p$,$pn$) (right) reactions at 400~MeV/nucleon. In all cases, the spectroscopic factor has been set to unity.}
\end{figure}

Once the double differential cross sections have been obtained,  the transverse and longitudinal momentum distributions are computed by means of Eqs.~(\ref{eq:dsdpx}) and (\ref{eq:dsdpz}), respectively. These quantities are shown in Fig.~\ref{c12p_dsdpx} (a) and (b), respectively. Note that the  longitudinal ($p_z$) momentum distribution has been calculated in the projectile rest-frame.  As in the case of knockout reactions between composite systems, the shape of these distributions is mostly determined by the wave function of the struck nucleon.  It is observed that the longitudinal momentum distribution exhibits some asymmetry and is somewhat shifted to negative values of  $p_z$. This is mainly a consequence of the negative $Q$-value of the reaction, which reduces the available kinetic energy in the final channel. This phenomenon has been analyzed in detail in a recent work in terms of the DWIA formalism \cite{Oga15}. The  calculations were repeated using the 2pF density of the $^{12}$C nucleus, but the results are almost identical, so they are not shown here.

Similar calculations for the $^{12}$C($p$,$pn$)$^{11}$C reaction have been performed. The ingredients are almost the same as in the $(p,2p)$ case, except for the fact that one needs  also the   $n$+$^{11}$C potential for the exit channel, which was also obtained  from the microscopic folding model, with the PH effective nucleon-nucleon interaction and the HF (Sk20) density of the $^{11}$C nucleus. The bound-state wave function was obtained with a Woods-Saxon well with diffuseness $a=0.7$~fm, radius $R_0=2.850$~fm (extracted from the HF calculation) and the depth adjusted to reproduce the experimental neutron separation energy ($S_n=18.72$~MeV).

\begin{figure}[t]
{\par\centering \resizebox*{0.45\textwidth}{!}
{\includegraphics{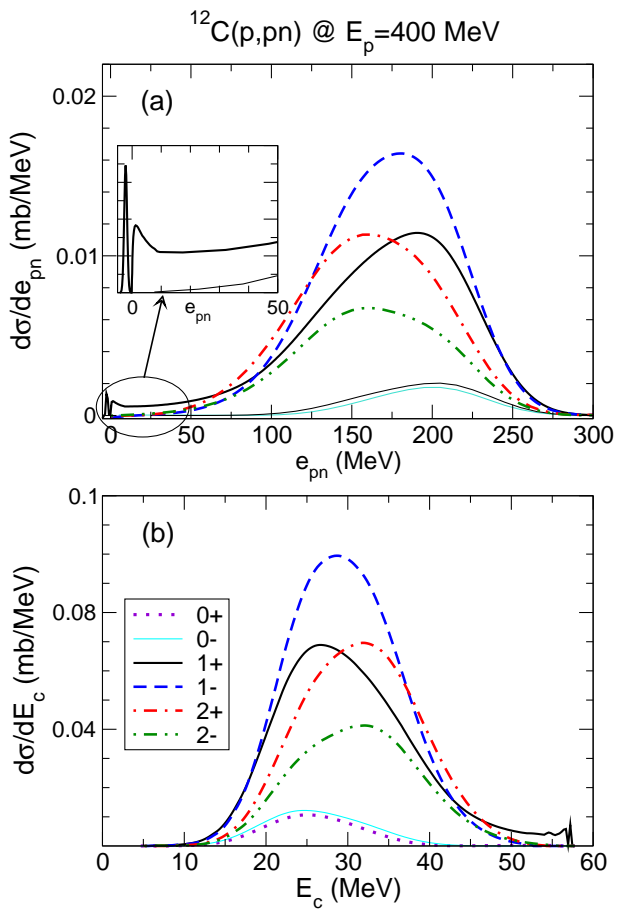}}\par}
\caption{\label{c12ppn_dsdex} (Color online) Top: relative energy distribution of the outgoing nucleons, following the process $^{12}$C($p$,$pn$), for some $j^\pi$ configurations of the outgoing $p$-$n$ pair. Bottom: corresponding energy distribution of the core, in the CM frame. See text for details.} 
\end{figure}

Figure \ref{c12ppn_dsdex} shows the calculated $(p,pn)$ cross section, as a function of the relative energy $p$-$n$ energy, $e_{pn}$, (top) and the core energy in the CM frame, $E_c$, (bottom). As in the  $(p,2p)$ case, the cross section is dominated by the $1^-$ partial wave of the $p+n$ system, although the contributions of the $1^+$, $2^+$ and $2^-$ waves are also sizable.  The $e_{pn}$ distribution corresponding to the  $1^+$ configuration exhibits a low-energy tail, with a sharp increase near $e_{pn}=0$. This peculiar behaviour is  a consequence of the final-state interaction (FSI) between the outgoing $p$ and $n$ nucleons. The contribution of the  $(p,d)$ channel, leading to the deuteron ground state, is also included, although it was found to be negligibly small at these relatively high energies. This is better seen in the inset, where the low-energy  region has been enlarged.
The FSI effect due to the $1^+$ wave is  also apparent in the core energy distribution (bottom panel), giving rise to a high energy tail. 

  The results for the transverse ($p_x$)  and longitudinal ($p_z$) momentum distributions are shown in Fig.~\ref{c12p_dsdpx} (c) and (d), respectively. As in the  $(p,2p)$ case,  the  struck neutron is also removed from the $p_{3/2}$ orbital, and so the shapes of the calculated momentum distributions turn out to be very similar in both cases. However, the magnitude of the $(p,pn)$ case is found to be $\sim$20\% larger. Using unit spectroscopic factors, one finds $\sigma(p,pn)=6.6$~mb and $\sigma(p,2p)=5.4$~mb and hence $\sigma(p,pn)/\sigma(p,pp)\simeq 1.2$, which is close to the free nucleon-nucleon cross sections at this energy,   
$\sigma_{pn}/\sigma_{pp} = 1.27$. This is consistent with a quasi-free scattering interpretation of this reaction. The departure from this free nucleon-nucleon cross sections can be attributed to the difference in separation energies, the Coulomb interaction and the slight differences in the matter densities (and hence in the nucleon-nucleus potentials).

\begin{figure}[t]
{\par\centering \resizebox*{0.45\textwidth}{!}
{\includegraphics{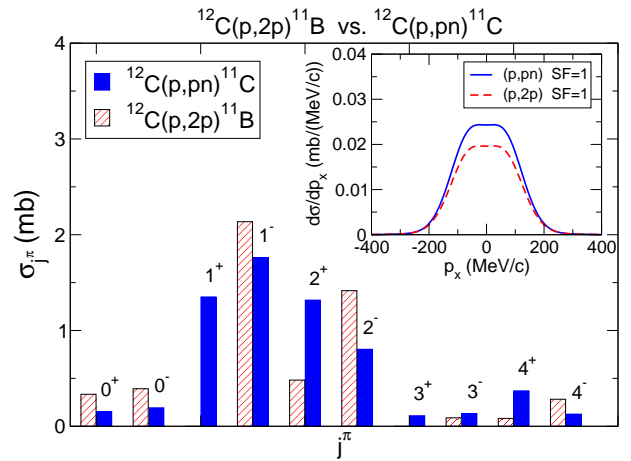}}\par}
\caption{\label{c12p_jpi} (Color online) Contribution of each angular momentum/parity ($j^{\pi}$) to the $^{12}$C($p$,$2p$)$^{11}$B and
$^{12}$C($p$,$pn$)$^{11}$C reactions at $E=400$~MeV/nucleon for a $p_{3/2}$ configuration and unit spectroscopic factor. The inset shows the corresponding transverse momentum distributions, summed over all the $j^\pi$ contributions.}
\end{figure}

To finish this section, the role of the different partial waves of the outgoing $p$-$N$ pair is discussed. This is shown in Fig.~\ref{c12p_jpi},  where the  histogram corresponds to the contribution of each $j^\pi$ to the $(p,pN)$ cross section, and the inset shows the total transverse ($p_x$) momentum distribution (summed in $j^\pi$). It is seen that most of the contribution comes from the $j=1,2$ waves. It is to be noted that  the pure $T=0$ configurations $1^+$ and $3^+$ are absent in the $(p,2p)$ case.  It is also seen that the contribution of $j\ge3$ is small and hence convergence of these observables requires only a small number partial waves. This is in contrast to the usual DWIA formulation, in which the final states are written in terms of the distorted waves for the outgoing nucleus, and hence a large number of partial waves is expected to be required in both waves to achieve convergence of inclusive observables.

\subsection{Application to $^{23}$O(\protect{$p$},\protect{$pn$})}
The  $^{23}$O(\protect{$p$},\protect{$pn$}) reaction at 450~MeV/u, populating bound states of the $^{22}$O nucleus, is considered now. In a simple mean-field picture, the single-particle configuration of the outermost neutrons of $^{23}$O is expected to be $\nu (1p_{1/2})^2(1d_{5/2})^6 (2s_{1/2})^1$.  Hartree-Fock calculations with the Skyrme Sk20 interaction yield the single-particle energies $\varepsilon(p_{1/2})$=-12.6~MeV, $\varepsilon(1d_{5/2})$=-6.1~MeV and $\varepsilon(2s_{1/2})$=-4.3~MeV. Since the neutron separation energy of $^{22}$O is  $S_n=6.85$~MeV, one would expect that only the removal of neutrons from the $2s_{1/2}$ and $1d_{5/2}$ orbitals of $^{23}$O  would lead to bound states of $^{22}$O. Removal from deeper orbits will lead to a residual $^{22}$O system with an excitation energy above its neutron separation threshold, which will  therefore decay into $^{21}$O. However, shell-model calculations, as those presented below, indicate that part of the $1p_{1/2}$ strength in the $^{23}$O nucleus corresponds to bound states of $^{22}$O. Consequently, this orbital has been also considered in the calculations of this section.

As in the previous examples, the required nucleon-nucleus potentials ($p$+$^{23}$O for the incident channel, and $p/n$+$^{22}$O for the exit channel) were obtained by folding the  PH NN effective interaction with Hartree-Fock (Sk20) ground state densities. Relativistic corrections were included according to the prescriptions discussed in Sec.~\ref{sec:relkin}, with the scaling factor of Eq.~(\ref{eq:gam2}) for the potentials. 
The effective separation energy for each single-particle configuration will depend on the considered state of $^{22}$O. In the present calculations, these states have been obtained from a  shell-model calculation, 
performed with  the code {\sc oxbash} \cite{oxbash} using the WBT effective interaction of Warburton and Brown \cite{WBT}. The results are summarized in Table \ref{tab:o23sf}. Guided by these results, for the $2s_{1/2}$ configuration we have considered the effective one-neutron separation energy corresponding to the transition to the $^{22}$O ground-state, whereas for the  $d_{5/2}$ and $1p_{1/2}$ configurations we have used the excitation energies of the $2^+_1$ and $1^-_1$ states, respectively. This gives the values 
$S_n=2.739$~MeV, $S^*_n=5.939$~MeV and $S^*_n=8.539$~MeV, for the $2s_{1/2}$, $d_{5/2}$ and $1p_{1/2}$ configurations, respectively. 

\begin{table}[b]
\caption{\label{tab:o23sf}%
Shell-model spectroscopic factors (SF) for   $^{23}$O$\rightarrow$$^{22}$O+$p$, corresponding to $^{22}$O bound states. The excitation energies and angular momentum-spin assignment are also those predicted by the shell-model calculation. See text for details.}
\begin{ruledtabular}
\begin{tabular}{lccc}
$E_\alpha$ (MeV)&
$I^\pi_\alpha$&
$n\ell j$ &
SF \\
\colrule
0 (g.s.) & $0^+_1$ & $2s_{1/2}$ & 0.80  \\
3.4  & $2^+_1$ & $1d_{5/2}$ &  2.08    \\
4.8  & $3^+_1$ & $1d_{5/2}$ &  3.08  \\
4.6  & $0^+_2$ & $2s_{1/2}$ &  0.12  \\
5.8  & $1^-_2$ & $1p_{1/2}$ &  0.76  \\   
6.1  & $0^-_2$ & $1p_{1/2}$ &  0.32  \\  
6.5  & $2^+_2$ & $1d_{5/2}$ &  0.24  \\
\end{tabular}
\end{ruledtabular}
\end{table}

\begin{figure}[t]
{\par\centering \resizebox*{0.45\textwidth}{!}
{\includegraphics{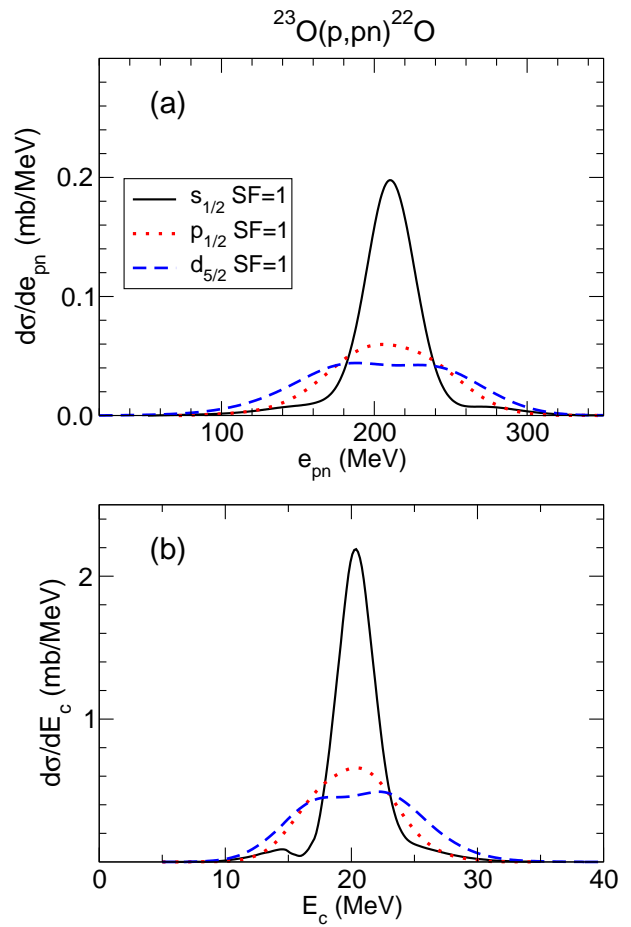}}\par}
\caption{\label{o23ppn_dsdex} (Color online) Top: Differential cross section as a function of the  outgoing proton-neutron relative energy for the reaction $^{23}$O($p$,$pn$) at  445~MeV/nucleon. Solid, dashed and dotted lines correspond to the removal from  $2s_{1/2}$, $1d_{5/2}$, and $1p_{3/2}$ configurations in $^{23}$O, assuming in all cases unit spectroscopic factor. Bottom: differential cross section as a function of the kinetic energy of the residual nucleus $^{22}$O in the CM frame. 
}
\end{figure}

Figure \ref{o23ppn_dsdex} shows the calculated $(p,pn)$ differential cross sections as a function of the $p$-$n$ relative energy (top panel) and the core energy in the CM frame (bottom). The solid, dashed and dotted lines correspond to the removal from  $2s_{1/2}$, $1d_{5/2}$, and $1p_{3/2}$ orbitals in $^{23}$O. The marked $\ell$ dependence is clearly seen, with the $\ell=0$ configuration giving a much narrower energy distribution. In this case, the $p$-$N$ pair will be detected with a relatively narrow distribution of energies centered at $\sim 210$~MeV, which corresponds to half of the incident beam energy, in accordance with a quasi-free NN scattering. For other values of $\ell$ the distribution is also approximately centered at this value, but with a larger dispersion.

\begin{figure}[t]
{\par\centering \resizebox*{0.45\textwidth}{!}
{\includegraphics{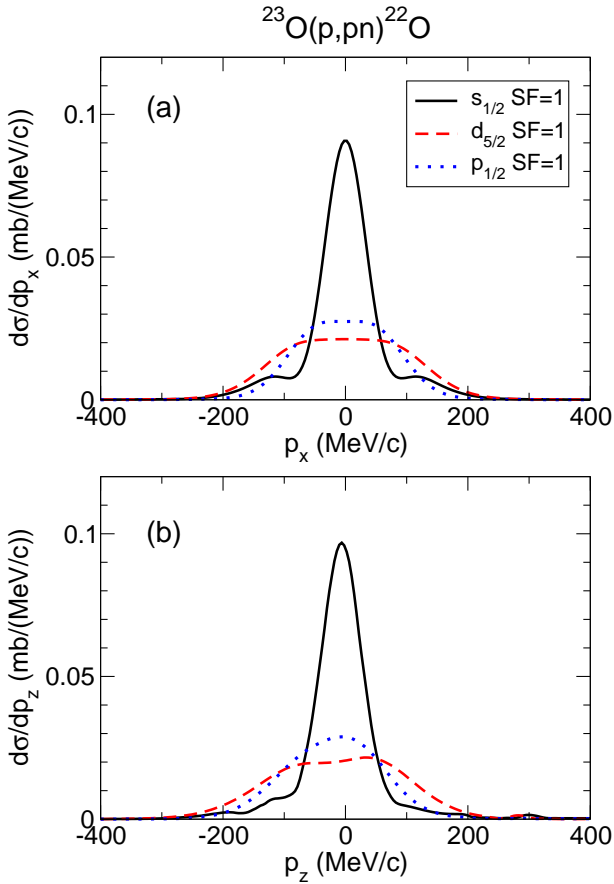}}\par}
\caption{\label{o23ppn_pxpz} (Color online) Transverse (top) and longitudinal (bottom) momentum distributions for the $^{22}$O redisual nucleus from the one-neutron removal of $^{23}$O at  445~MeV/nucleon. Solid, dashed and dotted lines correspond to the removal from  $2s_{1/2}$, $1d_{5/2}$, and $1p_{1/2}$ configurations in $^{23}$O, assuming in all cases unit spectroscopic factor.}
\end{figure}

The calculated momentum distributions, using unit spectroscopic factors, are displayed in  Fig.~\ref{o23ppn_pxpz}. 
The integrated $(p,pn)$ cross sections are $\sigma_{sp}=9.1$~mb, 6.1~mb and 5.7~mb, for the  $2s_{1/2}$, $1d_{5/2}$, and $1p_{1/2}$ orbitals, respectively, which might suggest a dominance of the $s_{1/2}$ contribution in the total $(p,pn)$ yield. However, these values are to be multiplied by the corresponding spectroscopic factors, which carry the information on the occupation number of each single-particle orbit. The calculated spectroscopic factors, using the shell-model calculations performed here, are listed in Table \ref{tab:o23sf}. 
The summed  spectroscopic for each configuration are: 0.92 ($2s_{1/2}$), 1.09 ($1p_{1/2}$) and 5.39 ($1d_{5/2}$). Compared with the mean-field  expectations (i.e. 1, 2 and 6 for the $2s_{1/2}$,  $1p_{1/2}$ and $1d_{5/2}$ orbitals), one sees that 
 removal from the  $2s_{1/2}$ and $1d_{5/2}$ orbitals will lead mostly to bound states of $^{22}$O, whereas about half of the $1p_{1/2}$ strength  lies in the continuum.

\begin{figure}[t]
{\par\centering \resizebox*{0.45\textwidth}{!}
{\includegraphics{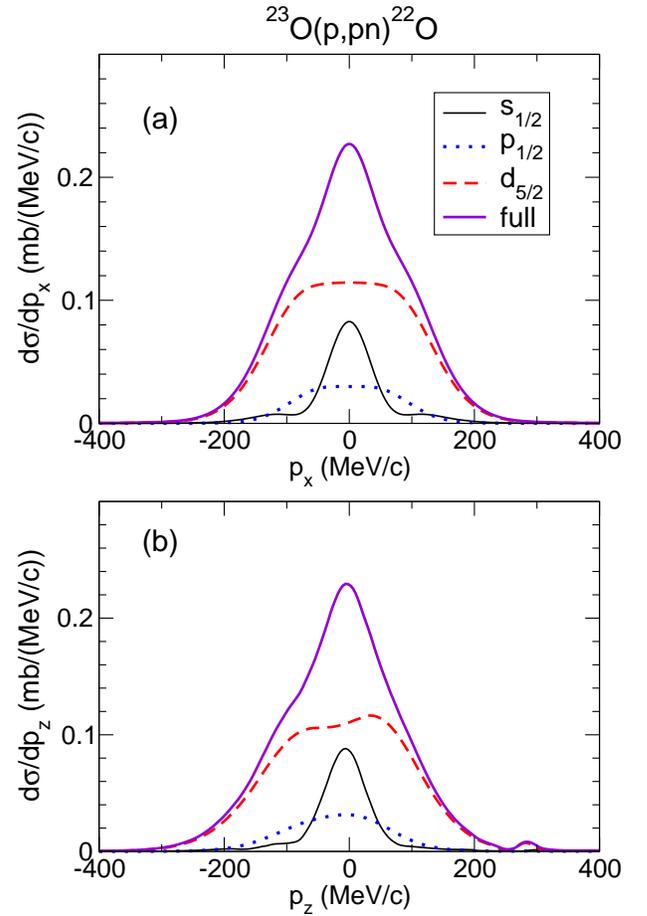}}\par}
\caption{\label{o23ppn_pzpz_inc} (Color online) Transverse (top) and longitudinal (bottom) momentum distributions for the $^{22}$O residual nucleus from the one-neutron removal of $^{23}$O at  445~MeV/nucleon. Solid, dashed and dotted lines correspond to the removal from  $2s_{1/2}$, $1d_{5/2}$, and $1p_{1/2}$ configurations, each of them multiplied by the summed spectroscopic strengths for these configurations corresponding to bound states of the $^{22}$O residual nucleus (see Table \ref{tab:o23sf}). 
}
\end{figure}

 In experimental conditions in which the state of the core is not identified (for example, without $\gamma$-ray coincidences), the measured momentum distributions will correspond to a superposition of all the bound states of $^{22}$O. The calculated inclusive spectrum is shown in Fig.~\ref{o23ppn_pzpz_inc}, where the thin solid, dotted and dashed lines are the $2s_{1/2}$, $1p_{1/2}$ and $1d_{5/2}$ contributions, weighted by the summed shell-model spectroscopic factors for each configuration. The thick solid line is the sum of the three contributions, i.e., the total inclusive spectrum. 
 Although dominated by the relatively broad $d_{5/2}$ component (indicative of the large occupation number of this orbital), a sharp peak corresponding to the narrower $2s_{1/2}$ orbital is also apparent. This marked $\ell$ dependence should permit the extraction of the $\ell=0$ contribution, even from inclusive measurements. More exclusive measurements, in which the $^{22}$O fragments could be detected in coincidence with deexciting $\gamma$ rays, are expected to be feasible in the near future, and will provide complementary information to that obtained with knockout measurements with heavier targets. A measurement of this kind has been made for the $^{12}$C($^{23}$O,$^{22}$O $\gamma$) reaction at 938~MeV/nucleon \cite{Cor04}. An important difference  between the two reactions is that, in the heavy-target knockout case, the inclusive cross section is dominated by the $s_{1/2}$ component ($\sim$60\%), whereas, according to the present ($p$,$pn$) calculations, the dominant contribution arises from the removal from the $d_{5/2}$ orbital ($\sim$69\%). This difference might be due to the larger penetrability of the proton target. In this respect, the ($p$,$pN$) measurements, being more sensitive to the deeper orbitals, can provide complementary information to that obtained from knockout experiments with  heavier targets.

\begin{figure}[t]
{\par\centering \resizebox*{0.45\textwidth}{!}
{\includegraphics{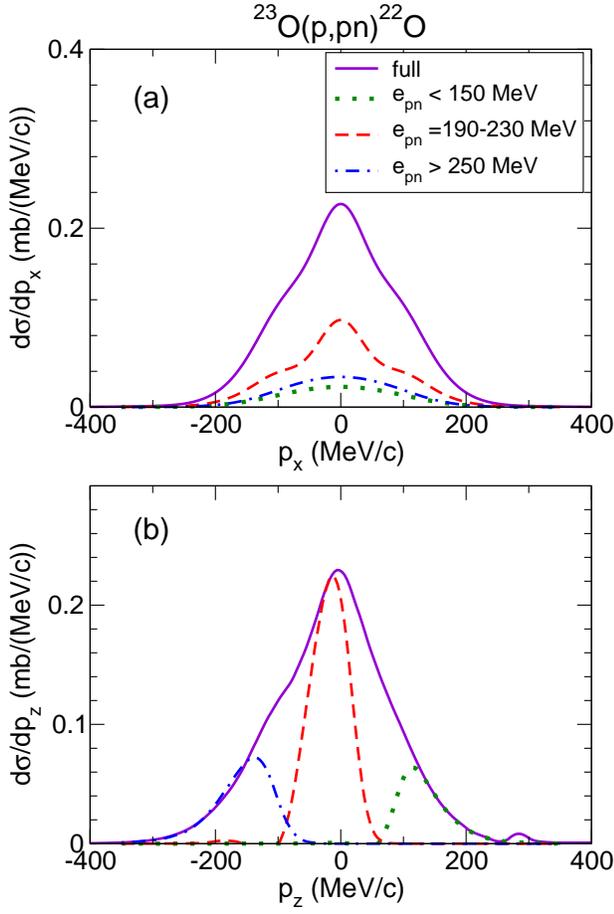}}\par}
\caption{\label{o23ppn_pzpz_ecut} (Color online) Transverse (top) and longitudinal (bottom) momentum distributions for the $^{22}$O residual nucleus from the one-neutron removal of $^{23}$O at  445~MeV/nucleon. The dotted, dashed and dot-dashed lines correspond to different cuts of the proton-neutron relative energy ($e_{pn}$). The solid lines correspond to the full distribution, integrated for all relative energies.  
}
\end{figure}

In the previous momentum distributions,  the full range of $p$-$N$ relative energies allowed by energy conservation was considered. In more exclusive measurements, in which the residual core is detected in coincidence with the outgoing nucleons, more detailed observables are possible. As an example, the momentum distributions constrained to specific intervals of the proton-neutron relative energy are considered now. This is illustrated in Fig.~\ref{o23ppn_pzpz_ecut}, where the dotted, dashed and dot-dashed lines are the summed $s_{1/2}$, $p_{1/2}$ and $d_{5/2}$ contributions, weighted by the shell-model spectroscopic factors, for the selected relative energy intervals $e_{pn}< 150$~MeV, $190 < e_{pn}< 230$~MeV and $e_{pn}> 250$~MeV. From Fig.~\ref{o23ppn_dsdex}, one expects that, for the $190 < e_{pn}< 230$~MeV cut, the contribution coming from the  $\ell=0$ orbital will be enhanced and this explains that this narrow contribution becomes more evident in the $p_x$ distribution as compared to the full spectrum. Conversely, for the $e_{pn}< 150$~MeV and $e_{pn}> 250$~MeV cuts, one is selecting mostly the $\ell=1$ and $\ell=2$ contributions and this explains that the corresponding $p_x$ distribution becomes broader. The impact of the different cuts is more dramatically evidenced in the longitudinal distribution (Fig.~\ref{o23ppn_pzpz_ecut}(b)). The smaller the final energy between the  $p$-$N$  pair, the larger the kinetic energy of the residual core, and the larger the values of $p_z$. 

\section{Discussion \label{sec:discus}}

The analysis of ($p$,$pN$) reactions has been usually done using the distorted-wave impulse approximation (DWIA) so it seems appropriate to discuss here the connection of this  formalism with the model proposed in this work. For that purpose, one may note that the solution of Eq.~(\ref{eqPsi}) can be formally written as
\begin{equation}
\label{eq:Psi3b_int}
\Psi^{3b(-)}_{f} = \Phi^{(-)}_f +  \frac{1}{E-i \epsilon - H }  V_i \Phi^{(-)}_f
\end{equation}
where $V_i \equiv V_{pN} + U_{pC} - U_{pA}$,  $H$ is the full Hamiltonian, and  $\Phi^{(-)}_f$ is a solution of 
\begin{equation}
\label{eqPhi}
[E^- - K_{r'}- K_{R'}  - U^\dagger_{pC} - U^\dagger_{NC} ] \Phi^{(-)}_{f} (\pvec{r}',\pvec{R}')=0
\end{equation}
Neglecting core recoil effects arising from the finite mass of $A$, the solution of this equation is given by the factorized form
\begin{equation}
\Phi^{(-)}_{f} = \chi^{(-)}_{p}(\br_p) \chi^{(-)}_{N}(\br_N) 
\end{equation}

 If the auxiliary potential $U_{pA}$ entering Eq.~(\ref{T3b_uaux}) is chosen adequately, for example, as the potential describing the elastic scattering of in the incident channel ($p$+$A$),   the remnant term $U_{pC} - U_{pA}$ will have a negligible contribution to the integral. Neglecting altogether this difference and inserting (\ref{eq:Psi3b_int})  into the transition amplitude (\ref{T3b_uaux}), one gets (with $S_{\alpha,\ell,j}$=1 for brevity)
\begin{align}
{\cal T}^{3b}_{if}(\alpha)  & =  \langle  \chi^{(-)}_{p}(\br_p) \chi^{(-)}_{N}(\br_N)  |  
t_{pN}  |\varphi^\alpha_{CA} \chi^{(+)}_{pA} \rangle ,
\label{T3b_Vpn}
\end{align}
where
\begin{align}
\label{tpn_3b}
t_{pN}(E) & = V_{pN}   \nonumber \\
          & + V_{pN} \frac{1}{E^+ -K_r - K_R - U_{pC} - U_{NC}-V_{pN}}  V_{pN} , 
\end{align}
which is the T-matrix describing the scattering of the incident proton with the struck nucleon in the presence of the interactions with the core. This is to be compared with the {\it free} $p$-$N$ transition amplitude, i.e., 
\begin{equation}
\label{eq:tfree}
t^f_{pN}(E_{pN})= V_{pN} + V_{pN} \frac{1}{E^{+}_{pN} -K_r -V_{pN}}  V_{pN} .
\end{equation}
They are formally related by
\begin{equation}
t_{pN}(E) = t^f_{pN}(E-K_R - U_{pC} - U_{NC}) .
\end{equation}
At sufficiently high incident energies, one may neglect  $U_{pC}$ and $U_{NC}$ in the propagator of Eq.~(\ref{T3b_Vpn}), thus resulting
\begin{align}
{\cal T}^{3b}_{if}(\alpha)  & =  \langle  \chi^{(-)}_{p}(\br_p) \chi^{(-)}_{N}(\br_N)  |  
t^f_{pN}(E-K_R)  |\varphi^\alpha_{CA} \chi^{(+)}_{pA} \rangle .  
\label{T3b_free}
\end{align}
This corresponds to the distorted-wave impulse approximation (DWIA). Since the T-matrix appearing in this expression still contains the kinetic energy operator $K_R$, a possible way of evaluating this expression is by inserting a complete set of eigenstates of the full kinetic energy operator $K_r + K_R$, which amounts at expanding the initial and final states in plane waves (see, e.g.~\cite{Tha67}).   A simpler approach would be to approximate $K_R \approx \frac{1}{2}E$, so that $t_{pN}(E) \approx t^f_{pN}(\frac{1}{2}E)$, which has been used, for example, in the context of proton inelastic scattering \cite{Ama80}.
Another common approximation in $(p,pN)$ analyses is the assumption that the T-matrix entering (\ref{T3b_free}) varies sufficiently slowly with momenta, and so its arguments may be replaced by their asymptotic values. In this case, the matrix elements of this T-matrix between these asymptotic momenta can be singled out from the integral, giving rise to a factorized expression for the scattering amplitude, and to  cross sections which are proportional to the free NN cross section. 

The main virtue of the DWIA method is that it treats the $p$-$N$ scattering to all orders, but at the cost of neglecting altogether the effect of the core in the transition operator. Note, however, that  distortion and absorption effects arising from the interaction of the core with the incoming proton and with the outgoing nucleons are taken into account in the distorted waves. In this sense, the full DWIA expression, Eq.~(\ref{T3b_free}) goes beyond the single-scattering approximation (SSA) representing the leading order of the Faddeev/AGS expansion \cite{AGS}. In fact, it has been recently shown in Ref.~\cite{Cres14} that, for the $^{12}$C($p$,$2p$)$^{11}$B reaction at 400~MeV/nucleon, this full DWIA expression reproduces very well  the  third-order AGS/Faddeev calculation.  
 However, at it was also pointed out in \cite{Cres14}, actual calculations based on the DWIA method employ additional approximations \cite{Cha77,Oga15}, namely, the NN t-matrix is evaluated using the asymptotic momenta,  and the off-shell NN t-matrix is replaced  by an average on-shell t-matrix in free space. With these approximations,  $|t^f_{pN}|^2$ is just the  free NN cross section multiplied by some constant.  Finally, an isotropic approximation is done to this NN cross section \cite{Aum13,Oga15}.  Clearly,  the validity of these approximations needs to be assessed for a correct interpretation of the experimental data.

To conclude this section, it is noted that the DWIA amplitude (\ref{T3b_free}) neglects final-state interactions (FSI) between the incident and struck nucleons and, in particular, the $(p,d)$ channel. This is in contrast to the formalism proposed here, in which these FSI effects are incorporated in the three-body final wave function. According to the calculations presented in the previous section, these FSI have been found to distort the energy and momentum distributions of the core. 
Moreover, although the contribution from the $(p,d)$ channel has been found to be negligible at the energy considered in these calculations,  they are expected to be more important at lower energies, and so the DWIA method might be inappropriate to analyze these situations.

\section{Summary and conclusions \label{sec:summary}}
A new method to describe nucleon knock-out reactions of the form $A(p,pN)C$ at intermediate energies has been presented and illustrated with specific examples.  The starting point of the method is the prior form transition amplitude for the effective three-body process $p+A \rightarrow p + N  + C$.  The three-body final wave function is approximated by a CDCC wave function, using an expansion of the three-body final states in  terms of eigenstates of $p$-$N$ Hamiltonian. To reduce the calculation to a numerically tractable form, the $p$-$N$ continuum is discretized in energy bins, and truncated in partial waves. 

The main features of the proposed method are the following: i) Final state interactions (FSI) between the outgoing nucleons are treated to all orders, ii) for the $(p,2p)$ case, the Pauli principle for the outgoing $p$+$N$ pair is also properly accounted for, using antisymmetrized proton-proton wave functions, iii) relativistic kinematics corrections are  taken into account in an approximate way, and iv) no impulse approximation is required in the derivation.
  
To reduce the ambiguity inherent to the choice of the optical potentials, required to reproduce the absorption suffered by the incoming proton and by the outgoing nucleons,  microscopic potentials generated with the energy- and density-dependent nucleon-nucleon effective interaction developed by the Paris-Hamburg group have been employed. The accuracy of this prescription at the energies of interest for current $(p,pN)$ studies, has been assessed against existing  $p$+$^{12}$C elastic data. The agreement with these data, once the suitable relativistic kinematics corrections are included in the Schr\"odinger equation, has been found to be rather satisfactory.

The proposed method is largely motivated by the current experiments being performed at intermediate energies (hundreds of MeV per nucleon). However, since the method does not make use of the impulse approximation, it may be applicable to  lower energies. Some knockout experiments on proton targets have  in fact been performed at energies of the order of 70-80~MeV/nucleon \cite{Kon09,Oza11,Sat14,Tsh14} and, thus, the comparison with the present theory could provide an assessment of the validity of this model at these energies. In particular, our formalism can include on the same footing the contribution from the pure three-body breakup and the $(p,d)$ channel, which might be important at these energies. Work in this direction is underway and the results will be published elsewhere.


Since the description of the final states is done in terms of the CM of the outgoing $p$+$N$ pair, the method does not provide in a straightforward way the exclusive cross section as a function of the angles or energies of these nucleons. On the other hand, it may provide a suitable framework for the analysis of inclusive reactions in which only the residual core is detected, as is the case in many experimental situations.  Ongoing and planned experiments of this kind will provide the community with a large body of experimental data, that will serve also as assessment of the validity of this and other theoretical models.

\begin{acknowledgments}
This work has been partially supported by project FPA2009-07653.  The author is deeply grateful to Ian Thompson for his valuable help in the numerical implementations of the {\sc fresco} code described in this work, to Kazuyuki Ogata for enlightening discussions regarding the relativistic corrections and to Mario G\'omez-Ramos and Joaqu\'in G\'omez-Camacho for many useful discussions and a critical reading on the manuscript.  
\end{acknowledgments}

\bibliographystyle{apsrev4-1}
\bibliography{qfs}

\end{document}